\documentclass[times,11pt]{article}

\usepackage{amsmath,amssymb,amsfonts}
\usepackage{algorithmic}
\usepackage{graphicx}
\usepackage{hyperref}
\usepackage{xcolor}
\usepackage[round,authoryear]{natbib}
\usepackage{pgfplots}
\usepackage{xspace}

\newcommand\minus{%
  \setbox0=\hbox{-}%
  \vcenter{%
    \hrule width\wd0 height \the\fontdimen8\textfont3%
  }%
}

\newcommand{\mi}{\minus}
\newcommand{\ear}{$\rho$\xspace}
\newcommand{\id}[1]{{\em #1}}

\renewcommand{\cite}{\citep}

\begin{document}

\title{A Novel Approach to Fairness in Automated Decision-Making using Affective Normalization}
\author{Jesse Hoey and Gabrielle Chan\\ 
David R. Cheriton School of Computer Science, University of Waterloo\\
200 University Ave. W.\\
Waterloo, Ontario, Canada, N2L2C9}

\maketitle

\begin{abstract}
  Any decision, such as one about who to hire, involves two components. First, a rational component, i.e., they have a good education, they speak clearly. Second, an affective component, based on observables such as visual features of race and gender, and possibly biased by stereotypes. Here we propose a method for measuring the affective, socially biased, component, thus enabling its removal. That is, given a decision-making process, these affective measurements remove the affective bias in the decision, rendering it fair across a set of categories defined by the method itself. We thus propose that this may solve three key problems in intersectional fairness: (1) the definition of categories over which fairness is a consideration; (2) an infinite regress into smaller and smaller groups; and (3) ensuring a fair distribution based on basic human rights or other prior information. The primary idea in this paper is that fairness biases can be measured using affective coherence, and that this can be used to normalize outcome mappings. We aim for this conceptual work to expose a novel method for handling fairness problems that uses emotional coherence as an independent measure of bias that goes beyond statistical parity. 
  
\end{abstract}

\section{Introduction}
\label{sec:introduction}
We view a {\em group} as a collection of $N$ autonomous, situated, agents. These agents have individual goals, agency (action space), and are able to record their history and make inferences based on it. Humans and some ``artificially intelligent'' computer programs are examples of agents. Taking a Bayesian view of probability as degrees of belief~\cite{Pearl88}, a {\em group belief state} is a probability distribution over a group's internal representations of their history, decentralized across all group members ($N$). On average, $1/N^{th}$ of it is handled by each member with only limited communication with all $N-1$ other members. The group belief state greatly depends on context; different groups, situations, and social networks within a group will imply different group belief states and different degrees of decentralization, possibly co-existing as humans simultaneously interact with multiple, hierarchical and overlapping social networks~\cite{RedheadPower2021}. For a given context, group and network, precision is a key element of this group belief state. As these group belief states contain beliefs about the future actions of all group members, how precise these distributions are will impact how predictable the social world is for each individual member.

At the same time, diversity in a group can be defined as the inverse of this predictability~\cite{Page2007}. A diverse group is less predictable as it has different agents with different perspectives, interpretations, strategies/heuristics and with different models of the world. An observed policy for an agent is a mapping from the real states of the world to the agent's actions. Agents sharing a policy will operate on the same plan, not necessarily act identically, but will be predictable. Agents with diverse policies will be less predictable because each agent's expectations of each other agent will be violated often. The other option for the diverse group to be predictable is for agents to {\em trust} each other, in that they are able to efficiently divide labor. Agent A trusts agent B will do task $T_B$ because they will do task $T_A$ and only $T_B+T_A$ gives a reward at all. The predictability is therefore over the {\em effects} of the other agent's actions, but is nevertheless present. 

Whether using authority or trust, equality between agents is ensured, but to a different degree. Even though homogeneity in the form of a unified policy implies predictability, it often fails to make outcomes from decision-making equally beneficial to everyone, and is historically enforced through strict central control to a varying degree. While such a policy, being too rigid, may be easily broken by any external shocks and is therefore highly fragile, an organizational culture of trust is much more flexible and strengthens as a result of external shocks. For example, two agents who trust each other do so more after they get through a difficult time together without either of them defecting.  As with many natural surviving processes, trust is antifragile and therefore crucial to consider in group dynamics~\cite{Taleb2012}. 

This implies that agents in a group are torn between conforming to the group and getting individual benefits. Each agent is trading off the predictability of their own future with respect to their preferences (implying individuality) with the predictability of the group (implying homogeneity). If an individual only optimizes the first (a classic example of the {\em homo economicus}) they would struggle and be ostracized because they would have enormous difficulty cooperating with others. On the other hand, if an agent only optimizes group conformity, they again would be worse off as they would not be satisfying their individual preferences. Conformity with the group, however, affords security, and so this option may be preferable to the first, even from an individual perspective.

This leads us to the central question of how best to coordinate a group of agents working together. As agents in a group would inevitably differ, they would likely diverge in opinion despite an initial shared interest. Some agents may be discriminated against as a result of the clash of opinions. What is the appropriate trade-off to conform to a discriminatory society while improving individual and or subgroup advantages? 
This paper explores the problem as the maintenance of a common consciousness~\cite{Durkheim1893} as the belief state of the group about its own future, including its own actions. The belief state of some (possibly distributed) entity can be thought of as how precise the future of that entity is with respect to its ecological niche, or econiche~\cite{BruchFeinberg2017}. 
Ecological fitness is often more simply described as the minimization of {\em surprise}: the better an agent is able to predict what will happen next, including its own actions, the fitter it will be~\cite{Friston_brain_2010}.



The mappings from individuals to group outcomes used in real human decision making consist of a combination of rational and affective components~\cite{HoeyMacKinnon2020}, corresponding to individual and group processes~\cite{Hoey2021b}. The rational component is how the outcomes are logically dependent on the relevant characteristics of the individuals only. The affective component is how the population {\em feels} about the individual-outcome mapping, which is heavily influenced by group membership, and which strongly impacts ethical and moral decision making~\cite{StarkHoey2020}. In this paper, we show a method for measuring the affective component, based on sociological social interactionist models~\cite{Heise2007}. Combined with a measure of the outcome mappings, we can perform an {\em affective normalization} on the outcome mappings to recover only the un-biased, rational component. 

Consider the following simple example, which we return to in Section~\ref{sec:sim}. Suppose a hiring decision is being made based in part on a mapping that assigns individuals with characteristic $A$ a better outcome ($O_A$) than those with characteristic $B$ (outcome $O_B$ with the value of $O_B$ less than the value of $O_A$ to the individual). For example, suppose B have some irrelevant property \ear, while $A$ do not, and $A$ get hired more than $B$ because of a bias against people with property \ear. Using this function will lead to hiring decisions that are Lipschitz unfair~\cite{Dwork2011}.\footnote{A Lipschitz unfair decision is one that unfairly disadvantages one group of people in that their distributions over outcomes are larger than expected.}  The degree of unfairness, however, can be estimated by asking a group of relevant people how they feel about each choice. Thus, they may feel comfortable with $A$ getting outcome $O_A$, but find that $B$ getting outcome $O_A$ makes them uncomfortable. Suppose they feel twice as comfortable about giving $O_A$ to $A$ than to $B$. This means that their estimate of the mapping function is likely also biased in the same way, and so those with characteristic $B$ should be getting roughly twice as much $O_A$ as they do under the original mapping function.

In the remainder of this paper, we first introduce a social-psychological theory of human interaction known as Affect Control Theory or ACT. This model will help us quantify the affective decision component, so we can remove it. We then introduce the concept of affective normalization, followed by two examples. We close with a more in-depth look at intersectional fairness in light of the affective normalization we are proposing.

\section{Methods}
\subsection{Affect Control Theory}
\label{sec:bayesact}
Affect Control Theory (ACT)~\cite{Heise2007, Heise2010} is a model of emotional coherence based in language that was founded on the control principle of~\citet{Powers1973}, which states something very reminiscent of the free energy principle: that people try to minimize incongruities by the deliberate act of controlling their perceptions. Heise~\cite{Heise2007} transposed this to the sentiment space of~\citet{Osgood1957}, imposed a denotative structure from symbolic interactionism~\cite{Mead1934}, and added affective dynamics~\cite{Gollob1974}. ACT is a computational model that has been used to predict classes of human behavior in a variety of settings~\cite{MacKinnonHeise2010}. ACT maintains a deterministic and static denotative model as an actor-behavior-object state (e.g. \id{manager hires student}), and an associated deterministic but dynamic connotative model: a dynamical system in Osgood's three-dimensional space of affective meaning mapping evaluation, potency and activity. This dynamical system represents values, or evaluative knowledge, which can be contrasted with declarative and procedural knowledge that are represented in the denotative model. The two models (denotative and connotative) are linked with a dictionary that maps from labels (e.g. \id{manager}) to sentiments (Osgood's three dimensions of emotional appraisal: evaluation - good vs. bad, potency - strong vs. weak and activity fast/loud vs. slow/quiet). These sentiments are elicited from a population of individuals using semantic differentials, which differ from Likert measures as they have opposing adjectives at each end. On a semantic differential, individuals rate a word, say \id{manager} on scales such as for evaluation with \id{good} at one end and \id{bad} at the other. The ratings are typically averaged across around $1000$ participants, but can be done across a much smaller group. The result for manager from the Indiana 2005 survey~\cite{Francis2006} is EPA:$1.0,1.6,1.3$.\footnote{The numerical scores range between $\mi4.3$ and $4.3$ for historical reasons} 

Emotional coherence in ACT is the difference between the sentiments elicited out of context, and the same sentiments elicited in a context given by an actor-behavior-object triple representing a situation. This difference (squared) is called {\em deflection}~\cite{Heise2007}, and measures how unlikely a given event is to occur. Thus, while \id{mother hugs child} is a low deflection (highly probable) event, \id{mother strikes child} is much higher deflection (less likely). The key insight of this paper is that these deflections can be used as an independent measure of bias in a population.

We can construct a set of actor-behavior-object events for a hiring decision with the deflections shown in Table~\ref{tab:deflects}.
 Also shown are the deflections for the behavior \id{fire-from-a-job} (EPA:$\mi1.1,1.5,0.4$)
\begin{table}[h]
\centering\begin{tabular}{ccccc}
 &\multicolumn{4}{c}{applicant}  \\
behavior  & \id{saleslady} & \id{student} & \id{delinquent} &  \id{criminal} \\ \hline
 \id{hire} &  1.1 & 1.1 & 3.2 & 4.1 \\
\id{fire} &  3.1 & 4.9 & 2.2 & 2.5 \\
\end{tabular}
\caption{\label{tab:deflects} Deflections for the event \id{manager} [\id{behavior}] \id{applicant}. A deflection close to $1.0$ is considered very low, a highly probable event. We use the Indiana 2005 dataset for these numbers~\cite{Francis2006}.}
\end{table}

Thus, someone labeled \id{manager}  would be more likely to \id{hire} (EPA:$1.7,1.9,1.1$) a \id{saleslady} (EPA:$0.6,\mi0.2,0.6$) or a \id{student} (EPA:$1.5,0.3,0.8$) than a \id{criminal} (EPA:$\mi2.4,\mi0.8,0.8$) or a \id{delinquent} (EPA: $\mi1.8,\mi0.8,0.4$), indicating a bias in the population against criminals and delinquents. This bias will also be part of estimates, by the same population, of how successful each of these hires is. That is, the same population will rate delinquents as having a lower chance of success.
More subtle differences, such as across gender, will yield smaller deflection differences. For example, the event \id{woman hire saleslady} has a deflection of 2.1, compared to 1.1 for \id{man hire saleslady.} By reversing the genders, we have uncovered a bias against women who hire people in the population surveyed in the Indiana 2005 dataset~\cite{Francis2006}.

The assignment of labels to individuals and behaviors is a key component of this analysis, which we have somewhat swept under the rug in the previous discussion. For example, the assignment of the label \id{manager} to someone may have to do with protected attributes such as race or gender. An applicant with some attribute (a particular gender or race, say) facing a hiring committee biased against that attribute may be labeled as a \id{delinquent} while an applicant without that characteristic may be labeled as a \id{saleslady.}
It is exactly this bias that we aim to remove by computing these deflections. The assignment of labels to specific groups, however, is information which needs to be carefully elicited. We have described potential methods for this elicitation in Section~\ref{sec:labeling}.

\subsection{Affective Normalization}
A general decision making problem can be stated as computing $P(outcome,a|e,\bar{e})$ where $a$ is the action to be taken, $e$ is some unprotected attributes we can measure, and $\bar{e}$ are some protected ones.\footnote{Protected features are ones on which a decision should not be based, such as race or gender. Unprotected features are ones that should impact the decision, such as education and experience.} A preference over outcomes then leads to the optimal decision to make. However, some parts of $\{e,\bar{e}\}$ may have both a rational impact and an affective impact on the estimate of this probability. We can measure the affective impact by simply taking the deflection of the ACT event $A=self, B=a, 0=\{e,\bar{e}\}$, and then computing
\begin{equation}
  P(a|e,\bar{e})\propto e^{-\hat{\alpha}\times deflection(\text{``self a} \{e,\bar{e}\}\text{''})},
  \label{eqn:boltz}
\end{equation}
where $\hat{\alpha}$ is an arbitrary scale factor.
Thus, the probability of the policy and outcome based on both sets of features (what would be measured in e.g. a decision based on both $e$ and $\bar{e}$), is 
\begin{equation}
 P(outcome, a| e,\bar{e}) = P(outcome|a,e,\bar{e})\times  P(a|e,\bar{e}).
\label{eqn:eqnaffdef}
\end{equation}
We can then assume that the rational outcome is not dependent on $\bar{e}$, the protected attributes, so we can recover the rational decision component as
\begin{equation}
  P(outcome | a,e) = \frac{P(outcome, a|e,\bar{e})}{P(a|e,\bar{e})}
\label{eqn:affnorm}
\end{equation}

That is, we can recover the rational component by measuring the probability of an outcome (e.g. by taking a poll) and also measuring people's emotional reaction to such actions and persons {\em in general}. The quotient of these two yields the remainder which is the rational decision policy.


\section{Exploratory Examples}
\label{sec:sim}
In this section, we go over two toy examples that demonstrate the versatility of the method. The first considers a biased hiring decision, while the second looks at a marketing choice.

\subsection{Hiring}
\label{sec:hire}
Suppose we have two attributes, gender, $r\in\{m,f\}$, and race, $r\in\{b,w\}$, which are protected in the sense that the decision making process should not depend on them. We also have one attribute, education $e\in\{m,b\}$, which we believe the decision making process should depend on (say this is a graduate or undergraduate degree). The population under study then have some (perhaps biased) $P'(success|e,g,r)$ for any setting of the three variables $e,g,r$ which they use for determining if someone should be hired. For each such setting, we can also construct an equivalent ACT event in actor-behavior-object space as ``person of gender $g$ and race $r$ with education $e$ is hired (by manager/self).'' Suppose that in some society, a racist bias, as well as a misogynist bias exists. Then simultaneously, one would expect that $P'(success|e=*,g=w,r=b)$, as measured in a population, would decrease, while in the equivalent ACT event, deflection would increase. Thus, a potential normalization of $P'(success|e,g,r)$ is possible, simply because deflection {\em is} the inverse of the probability of the event being observed. However, we can measure these two things using separate methods, allowing for the normalization. A direct elicitation of $P'(success|\cdot)$ (using a questionnaire or historical data of hiring practices) can be normalized by a deflection computed from sentiments about identities and behaviors elicited from the same population. If the deflection for the event was still low, although $P'(success|\cdot)$ was also low, then the normalized $P'(success|\cdot)$ would stay the same: there is no bias against hiring this person, so the low $P'(success|\cdot)$ must be ``real'' and this person should not be hired.

This means that a machine learning algorithm that learns $P'(success|e,g,r)$ may inherit a bias from the dataset on which it is trained. However, that exact same bias can be measured in the population who generated the dataset (the persons of different $e,g,r$ who were hired and either succeeded or were fired), and thus it can possibly be accounted for. The result is an unbiased machine learning algorithm that presents only the rational facts about success and failure based on unprotected attributes $e$. The machine learning algorithm can then provide an unbiased ranking of candidates, say. Hiring decisions can then be made as usual but decisions need to be justified with reference to the unbiased estimate. Suppose the algorithm places candidate $c_1$ (who has $e=m,r=b$) first, and candidate $c_2$ (who has $e=m,r=w$ second), although the estimated $P(success)$ rates are very similar. In this simple two-candidate example, $c_1$ would be hired by the algorithm over $c_2$, since it knows about the bias against $r=b$ and adjusts $P'(success|e=m,r=b)$ upwards. However, the algorithm only knows about $e$, while the committee knows as well about some other attribute, $s$, say the amount of community service done by the candidates (it could be more intangible than this). Given the algorithm has pointed to $c_1$, the committee will need to justify that $c_2$ gets the job because although $c_2$ has a lower $P(success)$, they also have $s$, which is very valuable to the employer, while $c_1$ does not. In order to hire $c_2$ without information about $s$, the committee would not have the justification and would side with the algorithm on $c_1$. 

\subsubsection{Affective Normalization}
Consider that $P(success|e,g,r)$ is in fact the product of two terms: an affective one that is based on deflection (i.e. ``I'd never hire a person of gender $g$ and race $r$''), and a rational one that is based on (bias-neutral) statistical occurrences in the past (i.e. ``of the past 10 hires, those of gender $g$ and race $r$ were successful 7 times).\footnote{The statistical occurrences themselves may or may not be the result of affective biases. Further, we are not considering other elements of the future beyond hiring. It may be that a certain subgroup performs poorly, not because of their inherent skills, but because of biases in the behaviors of their co-workers on the job (e.g. microaggressions).} That is, this rational term is difficult to measure because of how hard it is to separate out the affective biases from the rational data interpretation. However, using Equation~\ref{eqn:affnorm}, we can write:
\[ P(\text{\em success}|e) = \frac{P(\text{\em success} |e,g,r)}{e^{-\hat{\alpha}\times deflection(\text{``hire~e,g,r''})}}. \]
For some $e,g,r$ combination that is rated with a low success rate, it can be ``rescued'' by the deflection of hiring that $e,g,r$ combination as being higher. We are considering here that this deflection is strictly independent of the probability of success based on anything other than stereotypes about this combination of $e,g,r$.

To simplify this analysis, suppose that $g,r$ take on two values jointly of $w$ and $d$ (call this combined attribute $gr$). Let's imagine that a measure of success in an actual organization is that $P(success|e=m,gr=w)=0.9$ while $P(success|e=m,gr=d)=0.6$. Similarly, make $P(success|e=b,gr=w)=0.7$ while $P(success|e=b,gr=d)=0.3$. We can represent these in the following matrix format, in which columns are the $e$ attributes of $m$ and $b$, while the rows are the $gr$ attributes of $w$ and $d$. 
\[
\begin{array}{ccc}
  & m & b \\
w&   0.9 & 0.7 \\
d&   0.6 & 0.3 \\

  \end{array}
  \]

In this sample we can see what may be a direct bias toward people with $gr=w$. Now consider deflection, and that the deflections are biased against persons with $gr=d$ as follows. First, we have to assign identities to the different actors involved. Suppose we estimate that someone with $e=m,gr=w$ will be labeled as a \id{saleslady} and someone with $e=b,gr=w$ a \id{student}. However, due to a negative stereotype, persons with $e=m,gr=d$ are labeled as \id{criminals} and those with $e=b,gr=d$ as \id{delinquents}. We make these associations manually here in the context of this toy example, but in Section~\ref{sec:labeling} we show how they could be extracted automatically from existing text corpora. 
Thus, the deflections are those in Table~\ref{tab:deflects}, repeated here: 
\[
\begin{array}{ccc}
  & m & b\\
w&  1.1 & 1.1 \\
d&  4.1 & 3.2\\
  \end{array}
  \]  
  taking the exponent of the negative of this (setting $\hat{\alpha}=1.0$) gives something proportional to the probability of success:
\[
\begin{array}{ccc}
  & m & b\\
w&  0.33 & 0.33\\
d&  0.02 & 0.04 \\
  \end{array}
  \]  
  The probabilities of \id{failing} also must be estimated, so suppose they are simply the inverses of the ones above, although we lift this assumption at the end of the section.
  We end up with two arrays for success and failure:
\[
\begin{array}{c|c}
success & failure \\ \hline
\begin{array}{ccc}
  & m & b\\
w&  0.9/0.33 & 0.7/0.33 \\
d&  0.6/0.02 & 0.3/0.04 \\
  \end{array}
&
\begin{array}{ccc}
  & m & b\\
w&  0.1/0.67 & 0.3/0.67 \\
d&  0.4/0.98 & 0.7/0.96 \\
  \end{array}
  \end{array}
  \]  
which, if normalized, gives us the final probability of success, which may be compared to the original matrix estimated from the biased data, also shown here for clarity
\[
\begin{array}{c|c}
\text{normalized~} P(\text{success}) & \text{original~} P(\text{success}) \\ \hline
 \begin{array}{ccc}
  & m & b\\
w&  0.95 & 0.82 \\
d&  0.99 & 0.91 \\
  \end{array}
&
\begin{array}{ccc}
  & m & b \\
w&   0.9 & 0.7 \\
d&   0.6 & 0.3 \\
  \end{array} 
  \end{array}
  \]  
We see that the success rate estimates have increased for the group with $gr=d$, and only somewhat increased (15\%) for the group with $gr=w,e=b$. The analysis above used deflection estimates that include an arbitrary scaling factor (to go from deflections to probabilities using equation~\ref{eqn:boltz}), that was set to $\hat{\alpha}=1.0$ and ended up with the reasonably equitable success rate estimate as shown above. 
  While theoretical estimates of this normalization factor may be possible,
  we show here a few other values for the scaling factor to give an idea of the tradeoff involved, as shown here:
   \[
   \begin{array}{c|c}
   \hat{\alpha}=0.5 & \hat{\alpha}=1.3 \\ \hline
\begin{array}{ccc}
  & m & b \\
w&   0.87 & 0.63 \\
d&   0.91 & 0.63 \\
  \end{array}
   &
\begin{array}{ccc}
  & m & b \\
w&   0.97 & 0.88 \\
d&   0.997 & 0.96 \\
  \end{array}
  \end{array}
  \]  
For $\hat{\alpha}=0.5$, the solution is still quite equitable, but less so that the $\hat{\alpha}=1.0$ solution. Increasing to $\hat{\alpha}=1.3$ gives a solution that is about as equitable as the $\hat{\alpha}=1.0$ solution, although overall more optimistic. 
   

  Looking back at the Lipschitz condition described earlier, the similarity metric in this simple case could be that persons with the same education level $e$ should be considered to be equal. The $[\text{\em rational}]$ component then must end up looking like
  \[
\begin{array}{ccc}
  & m & b \\
w&   p_m & p_b \\
d&   p_m & p_b \\

  \end{array}
  \]  
where $p_m$ and $p_b$ are the probability of success for candidates with $e=m$ and $e=b$, respectively. We can certainly see that the normalized distribution is much more fair than the original one in terms of this definition of similarity. For both levels of $e$, the $P(success)$ values have drawn closer together, meaning that the Lipschitz condition is more well satisfied. We do not consider population sizes in this example, so the mapping above is also marginally fair across the protected attribute.

 \subsubsection{Revised Normalization}
\begin{figure}
  \centering
  \begin{tabular}{ccc}
    \includegraphics[width=0.33\textwidth]{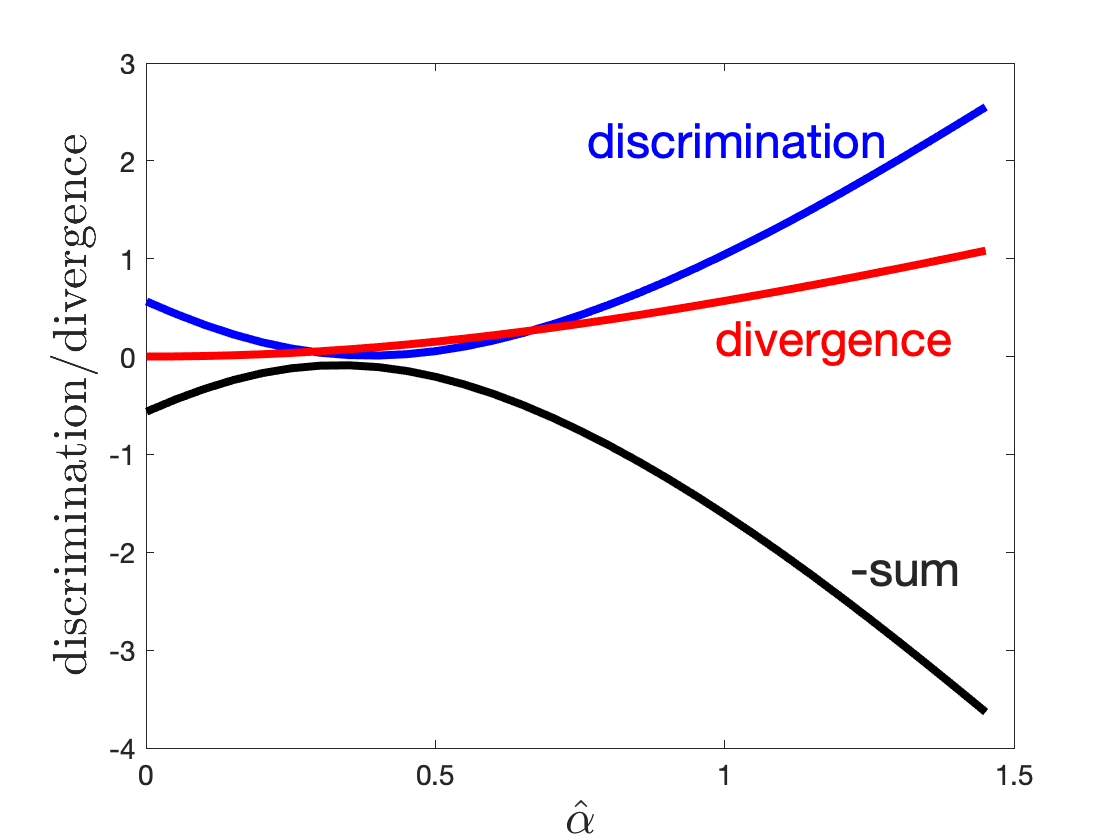} & 
    \includegraphics[width=0.33\textwidth]{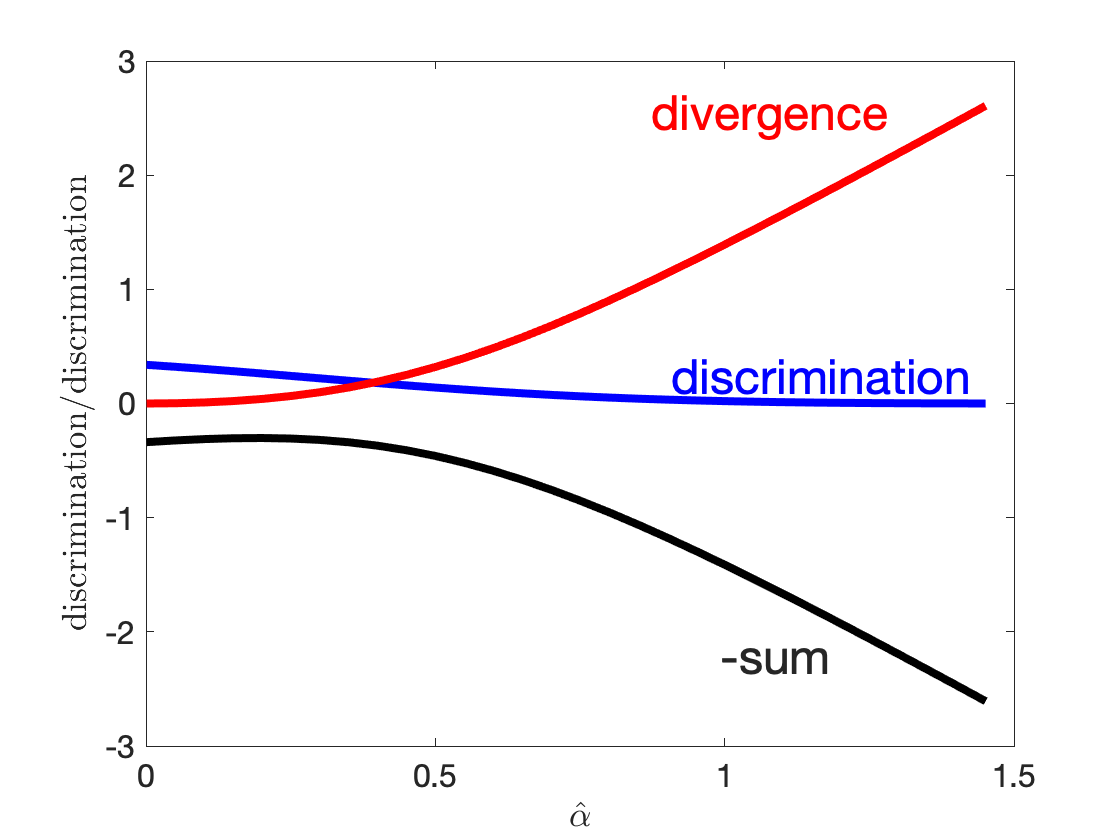} & 
    \includegraphics[width=0.33\textwidth]{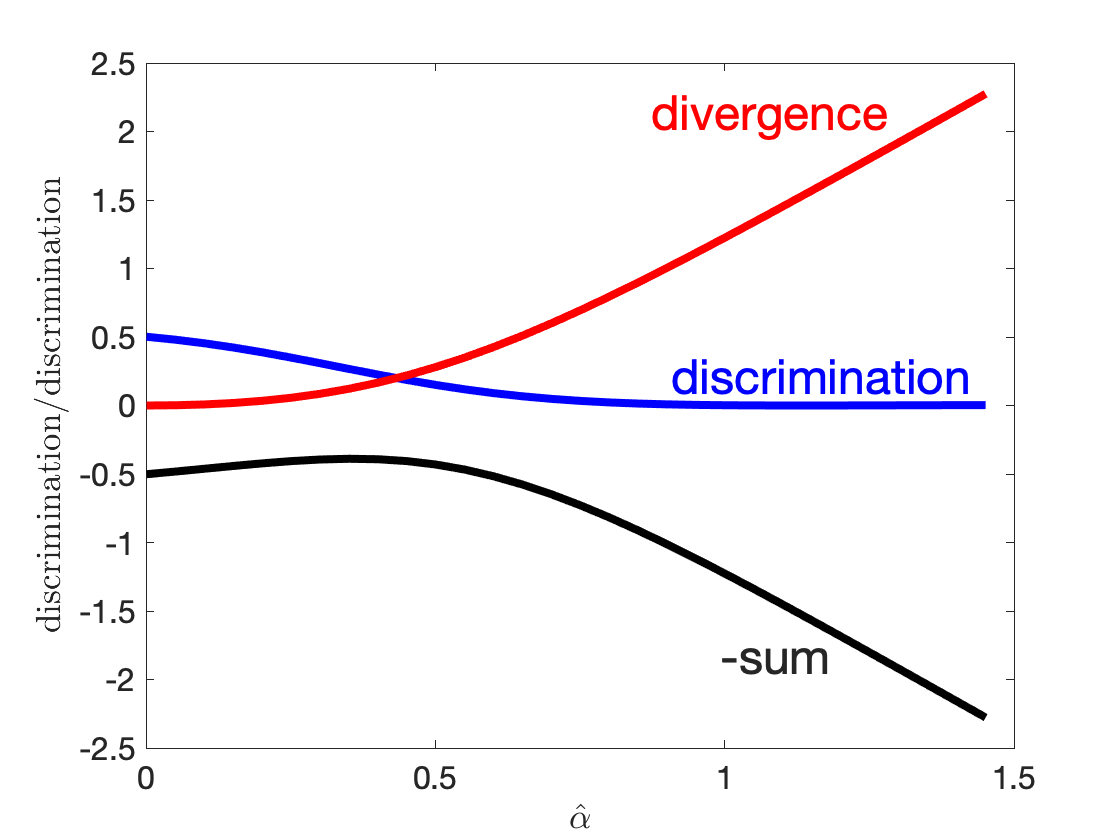} \\
    (a) & (b) & (c) \\
    \end{tabular}
    \caption{\label{fig:alphacurves} Discrimination (blue) and model divergences (red) combine linearly to give (black) an best estimate of where both divergence and discrimination are minimized. The model predictions trade off equity (Lipschitz) with accuracy (how well they optimize the employer's loss function), but other weightings may also be possible. (a) hiring, Section~\ref{sec:hire}, max at $\hat{\alpha}=0.35$ (b) biased marketing, Section~\ref{sec:market}, max $\hat{\alpha}=0.2$ (c) biased marketing, marginal outcome mapping in blue, $\hat{\alpha}=0.4$.} 
\end{figure}

We now return to the estimation of the $P(failure)$ using the deflection, which above we simply assumed was $1-P(success)$. This assumption is not correct because the emotional impact of hiring someone is not the direct inverse of the emotional impact of not hiring someone. That is, we might instead use the event \id{manager fire O}, where O is the applicant. 
These deflections are also shown in Table~\ref{tab:deflects}. Doing this, and repeating the process for a range of $\hat{\alpha}$ gives a set of unbiased probability matrices that we can compare to the original $P({\text success})$. To make the comparison, we look at (1) how inequitable it is across the protected attribute, 
show as the KL-divergence of the distribution across the protected attribute, averaged over the unprotected one, shown in blue in Figure~\ref{fig:alphacurves}(a), and (2) how different it is from the original, shown as KL divergences between the normalized and original distributions, averaged across all four conditions, shown in red in Figure~\ref{fig:alphacurves}(a). 

If we simply sum these,\footnote{how to balance these two is an empirical question based on the domain. Other notions of fairness beyond statistical parity, such as differential~\cite{foulds2019intersectional} can also be used as measures of equity.} we get a maximum at $\hat{\alpha}=0.35$ (determined visually), which corresponds to the following unbiased probability matrix:
\[
\begin{array}{ccc}
  & m & b\\
w&  0.81 & 0.38 \\
d&  0.73 & 0.38\\
  \end{array}
  \]   
So we can see the method is favoring equitable distributions, in particular over the $e=b$ class. Another way to interpret the blue curve in Figure~\ref{fig:alphacurves} is as a measure of how far the distribution is from satisfying the Lipschitz condition above. 
After the inflexion point at $\hat{\alpha}=0.35$ in Figure~\ref{fig:alphacurves}, the closest is achieved, and this is close to zero, after which the originally disfavored group starts to gain disportional advantage. 

The use of KL-divergences in Figure~\ref{fig:alphacurves} is a modeling choice we have made. The use of a variation norm gives a slightly different optimal solution at $\hat{\alpha}=0.6$ of 
\[
\begin{array}{ccc}
  & m & b\\
w&  0.75 & 0.26 \\
d&  0.78 & 0.41\\
  \end{array}
  \]

\subsection{Marketing}
\label{sec:market}
Consider a similar problem of a marketing choice from \citet{Dwork2011} where two ads, $a_0$ and $a_1$ are shown to people based on their credit score (group $G_0$ with bad credit, $G_1$ with good credit), with the terms of $a_1$ being superior to those of $a_0$. People are also divided into two groups according to some protected attribute: $T$ (majority) and $S$ (minority). Suppose first that, unlike in~\cite{Dwork2011}, the vendor has access to the protected attribute, so he can tell the person's credit score and whether they are in group $S$ or $T$. We will also start by assuming, like in the previous example, the populations are the same size, but the vendor has a bias against members of one group, $S\cap G_0$ giving the probability of showing $a_1$:
  \[
\begin{array}{ccc}
  & G_1 & G_0 \\
T &   0.9 & 0.7 \\
S &   0.9 & 0.3 \\
  \end{array}
\]

Due to the same bias, members of $G_1$ are labeled as \id{workers} but members of $G_0$ are labeled \id{victims} if they are in $T$, but \id{loafers} if they are in $S$. The deflections for showing $a_1$ are computed as the events \id{salesman sell something to $X$} while those for showing $a_0$ are \id{salesman cheat $X$}, where $X$ are the ids above (\id{workers}, \id{victims} and \id{loafers}). The deflections for showing $a_1$ and $a_0$ to each of the four groups is:
\[
\begin{array}{c|c}
a_1 & a_0 \\ \hline
\begin{array}{ccc}
  & G_1 & G_0\\
T&  1.5  & 3.1\\
S&  1.5 & 3.6 \\
  \end{array}
&
\begin{array}{ccc}
  & G_1 & G_0\\
T&  6.6 & 5.6 \\
S&  6.6 & 5.0 \\
  \end{array}
  \end{array}
  \]  
  which results in the plot show in Figure~\ref{fig:alphacurves}(b), and the normalized outcome mapping using the maximum $\hat{\alpha}=0.2$:
  \[
\begin{array}{ccc}
  & G_1 & G_0 \\
T &   0.76 & 0.58 \\
S &   0.76 & 0.25 \\
  \end{array}
\]
We observe that the probabilities have shrunk, meaning $a_1$ is shown overall less often, but the parity for $G_0$ is increased.

Let us now lift the assumption that the vendor can tell the difference between $S$ and $T$. The outcome mapping is already fair according to the criteria above, but the marginal probability of $S$ seeing $a_1$ overall is greatly reduced because the group $S\cap G_1$ is very small. In this case, the vendor can reduce the probability of showing $a_1$ to $G_0$, thereby unfairly disadvantaging $S$. The outcome mapping now does not differentiate between $S$ and $T$:
  \[
\begin{array}{ccc}
  & G_1 & G_1 \\
T &   0.9 & 0.1 \\
S &   0.9 & 0.1 \\
  \end{array}
  \]  
  However, $S$ is a minority with few people in $G_1$: assume the populations are $|S\cap G_1|=50$, $|S\cap G_0|=|T\cap G_0|=200$, and $|T\cap G_1|=800$. The marginal outcome mapping is the likelihood of any given member of $[S,T]$ seeing $a_1$.

  While in the hiring example and the first part of this marketing example, the biases were those of the hiring committee or vendor, in this case, it is the biases of the population that are important as we are attempting to make the distribution marginally fair (so persons in $S$ overall have a more similar chance of seeing $a_1$ to persons in $T$). Clearly, this cannot be a bias from the vendor, as they cannot tell the difference (they cannot tailor their ad presentations according to $S/T$, but they do know the population sizes). Therefore, we keep the deflections the same as in the first part, as they are measures of the wider population's view of the marketing choice. Figure~\ref{fig:alphacurves}(c) shows the result, with the blue line now indicating the discrimination across $S$ and $T$ overall, taking population size into account. The optimal $\hat{\alpha}=0.4$ yields the outcome mapping:
  \[
\begin{array}{ccc}
  & G_1 & G_0 \\
T &   0.54 & 0.04 \\
S &   0.54 & 0.06 \\
  \end{array}
\]
which has a marginal distribution of $[0.44,0.15]$ over classes $[T,S]$ compared to the original mapping which gives $[0.74,0.26]$. Thus, $a_1$ is being shown much less often, but the marginal difference is only $.29$ instead of $.48$. If we ignore the vendors outcome divergence (essentially ignoring their utility function), we set $\hat{\alpha}$ to be very large, which gives the degenerate solution of never showing $a_1$ to anyone. 
  
\section{Intersectionality}
\label{sec:intersectionality}
Many modern notions of fairness (e.g. comparing differential or subgroup distributions) don't necessarily satisfy (1) definitions of categories, and a mechanism for adding or changing categories e.g. who is a marginalized group and what if a new marginalized group appears or one disappears?; (2) an appropriate resolution of these categories of people, to avoid an infinite regress; and (3) the focus on equal distribution~\cite{Kong2021}.

The first two considerations arise in the classic example of hiring decisions that explicitly encourage black or female hires in the presence of racial and gender biases in hiring. One may implement a policy that favors persons of color, and another that favors women, but together these two policies may be discriminatory to black women. 
To see why, notice that performance (increased hiring) according to one policy may lead to more black people being hired, but due to the gender bias, in fact more black men will be hired than black women. Similarly, acting by the other policy may lead to more women being hired, but due to the race bias, in fact more white women will be hired than black women. Thus, black women are unfairly doubly disadvantaged by the implementation of these two policies, which instead require an intersectional policy of hiring specifically black women.

The third issue with intersectionality is equality of distribution, which is the inverse problem to the one stated in the last paragraph. That is, the differentiation of the sub-groups is there for a reason: to help make distribution of resources more equitable. Therefore, a balance must be struck. 

\subsection{ACT model}
Affective normalization and ACT may provide a computational model for each of these issues, leading to tools to investigate possible solutions. ACT gives a way of overcoming the category labeling problem by ``shaping the meanings of words themselves''~\cite{Wellens2008}. That is, if two sub-groups are placed in the same category affectively in the population, then it means they are indistinguishable in terms of how the same population would make a rational decision concerning those sub-groups. That is, the differentiation of the subgroups is not {\em affectively} significant, and thus a measurement of this decision would be fair to these subgroups. The category labeling problem then becomes a population measurement problem (of their affective feelings about events and subgroups in general). 

If there is no bias, deflections will be the same for both groups and the outcome variance will be revealed as valid.
Suppose we were evaluating applicants for a job as a dog trainer, and our algorithm is not distributively fair because does not hire enough canophobes (people who fear dogs), all else being equal. However, canophobia being anti-required for the job, sentiment towards ``canophobes not hired by dog training service'' would have the same low deflection (would be considered socially normal) as for any applicant. Thus, differentiating canophobes from non-canophobes is still necessary and would play into hiring decisions.  

The infinite regress problem in intersectionality occurs because regardless of the bias in a machine learning algorithm based on some categorization of people into groups, some dataset can always be created with more groups (with a finer grained categorization) which exposes some unfairness. While all would agree this regress must be stopped at some point (if not for political stability, at least for computational tractability), no one knows what that stopping point should be. ACT defines the stopping point in terms of emotional coherence. It caps the regress when there ceases to be any incoherence, which means there is little affective bias in the population under study. For a subgroup to be recognized and promoted by an artificially modified outcome mapping, a measurement of the population can reveal the emotional bias against this subgroup, which can flagged to be put under consideration as not necessary for the algorithm. If no such bias in the population exists, the method cannot suggest that the bias be removed, and the category may be flagged as one that may be put under consideration as necessary. 

Finally, the third problem is directly addressed by the normalization we propose, as this is aimed at ensuring equality. It is also addressed with prior information based on constitutional or human rights (or anything else) which can be directly inserted into the model and which fundamentally biases what choices can be made. 

The example in the last section showed how an affective bias could be accounted for in an outcome mapping. When a group is disfavored intersectionally, this is typically revealed by a subgroup raising its voice. Deflection measurements of the population under study to such a subgroup raising its voice might reveal that, indeed, there is a bias that can be addressed. Should such measurements reveal no such bias, then the outcome mapping may be left undisturbed.  Following the example from Section~\ref{sec:sim}, if the group $gr=d$ differentiates into $gr=da$ and $gr=db$, and $db$ claims a biased outcome mapping is being used in their disfavor, the bias towards the subgroups in the population under study can be compared to those towards the original super-group. Should there be a substantial difference, then the two factors intersect affectively, and the algorithm would recommend $db$'s outcome mapping be improved. If not, then the regress is capped, and no further differentiation is possible on this branch.

\subsection{Category Labeling and Word Embeddings}
\label{sec:labeling}
The method we propose involves measuring people's attitudes and sentiments towards events that are ongoing, something that is not trivial, and may require significant further advances to be able to do correctly. Much ongoing work in affective computing is aimed at unobtrusively measuring emotional states, such as depression~\cite{Lewis2021}. 
However, there is a long way to go from there to measures of sentiments about language as used by ACT. While semantic differentials may be used to assess these sentiments, this is time consuming and slow. Instead, data mining of social media may be another way to make progress on this issue~\cite{kouloumpis2011}. Here discuss more fully how to get from raw text corpora (e.g. company meeting minutes and emails) to (1) narrative descriptions of events as above and (2) EPA ratings for these narratives. (1) will be harder to obtain as it involves gaining an understanding of how people frame their communities of practice. Attempts to solve these problems automatically have been recently published~\cite{JosephMorgan2021,VanLoon2022}.

The first problem is to convert the situation being modeled to a narrative in the ACT grammar of Actor-Behavior-Object. In the exploratory example above, we selected identities from an existing dictionary~\cite{Francis2006} that matched the situation. However, the situation was constructed to explore the method, and so to apply the method realistically, methods must be sought to automate this process. For example, \citet{JosephMorgan2021} use the concept of a {\em Latent Coginitive Social Space} which combines social cues including sentiment, socio-demographics, and institutional affiliations to estimate identities automatically from text corpora and descriptions. 

The second problem is that we are limited by publicly available sentiment dictionaries to form label distributions in a sentiment space as in ACT. In other words, we need to find ratings given by the group we are interested in for the specific words involved. In the manager example, we have to elicit ratings from those doing the hiring. To apply this method more effectively and fairly, we need to discover automated approaches to assign labels to specific groups based on both algorithmic and human-influenced decision-making. To address this bias, word embeddings may present an opportunity to measure people’s attitudes and sentiments towards events that are ongoing, to scale.

Word embeddings are a product of natural language processing technologies and create vector relationships of words. From health inequities to refugee policies, word embedding maps have already proven to be an excellent decision-making tool by uncovering key connections between current system failures and potential solutions~\cite{Turan2019}. To address the gaps in underreported bias types~\cite{Rozado2020}, there exists a need to study underrepresented groups by deliberately seeking material written by members of the groups and encoding such research into public-facing word association libraries~\cite{lepori2020unequal,Nichols2019}. Efforts are underway to use word embeddings to estimate sentiments in EPA space from existing texts shared by members of the group of interest~\cite{VanLoon2022}, and to estimate cultural structures more generally~\cite{kozlowski2019geometry}.  \citet{VanLoon2022} generate a standard word embedding space and then locate the dimensions of E,P,A within it, leading to a sentiment score for each word in the space. In general, data mining of social media promises to tackle this issue~\cite{Joseph2016,kouloumpis2011}. Ultimately, mapping values of protected attributes to sentiments through word embeddings may serve as a starting point for the proposed model to iterate on human affect dynamics and predict agent behavior in a group setting.

\begin{figure}
    \centering
    \includegraphics[width=0.7\textwidth]{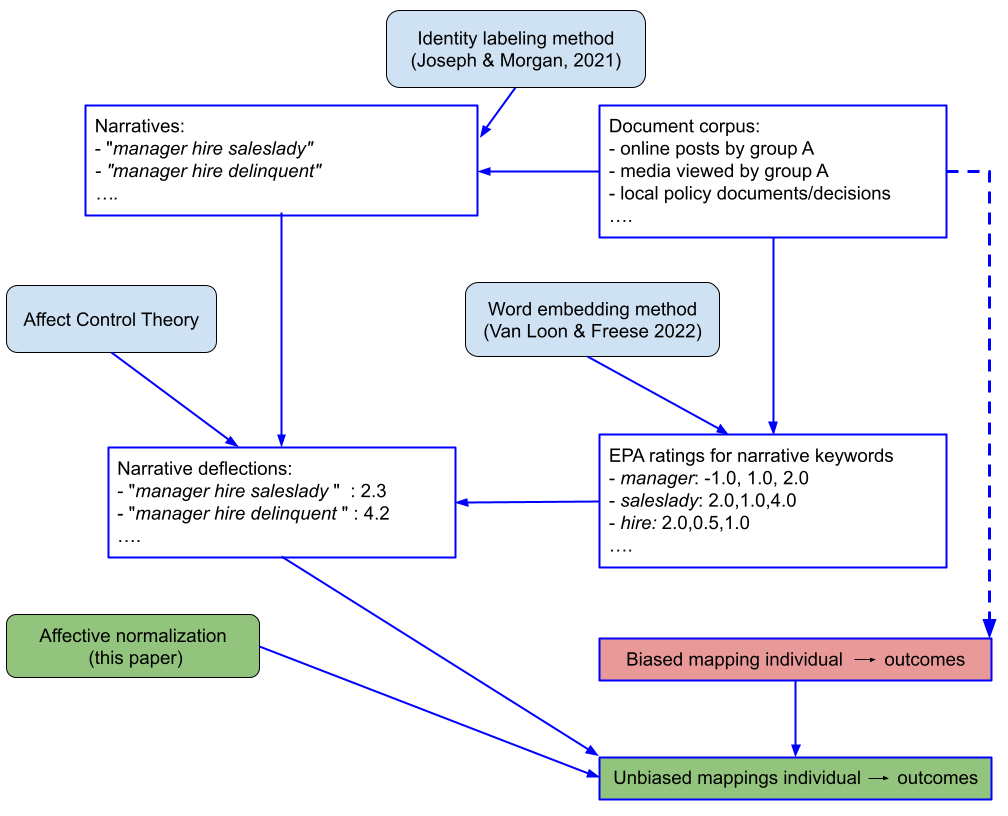}
    \caption{\label{fig:schematic} Schematic of the proposed method for affective normalization of outcome mappings. The part shaded in grey is not fully implemented yet, but is discussed in Section~\ref{sec:labeling}.}
\end{figure}

Figure~\ref{fig:schematic} shows the overall process that we are proposing. A document corpus is used to extract narratives and biased outcome mappings (these could also be elicited in some other way, e.g. using surveys). The narratives are converted to A-B-O triples using word embeddings. Deflections for these entities are computed using Affect Control Theory, and the affective normalization process based on these deflections is applied to the biased outcome mapping to yield an un-biased version. 

\section{Discussion}
The motivation for the ideas in this paper arises from the now famous example of racial bias is the commercial gender classification failure in facial detection software that had an error rate of up to 34.7 percent for dark-skinned women and only 0.8 percent for white men~\cite{Buolamwini2018GenderSI}. This is due to the original algorithm developers' bias when creating the training dataset for the software and not including sufficient examples of non-white, female subjects. If such decisions were ``checked'' in some way according to the affective normalization scheme we propose, this bias would be immediately revealed. 

To see the broader importance of these biases, consider that disturbances of social order, historically speaking, are often led by minority groups seeking change to the status quo due to perceived discrimination. From current Indigenous reclamation initiatives in North America to the decades-long women’s suffrage movement beginning in the mid-19th century, our history is littered with examples of fighting for equality as people rallied together to bring attention and enact positive change on systemic biases. However, so-called victories, such as the 1918 Canadian federal decision to allow “women” over the age of 21 to vote, often left out subgroups of the discriminated faction. In the aforementioned example, Asian and Aboriginal women, both classified equally “female” as white women, were not allowed to vote until respectively 1948 and 1960, bringing to attention the effects of intersectionality~\cite{Scotti2017}. It is reasonable to assume that white (usually property-owning) women gained an advantage over women of other racial origins or lower economic status due to racist and classist beliefs, and it can even be argued that white women’s progress in gaining the right to vote somewhat hindered others’ self-advocacy efforts~\cite{Grimshaw1999}. In order to understand this process, and to ensure future such processes take fairness considerations into account, the computational model we propose may help in quantifying one particular component, making it removable in a simplified way.

The proposed method is not a mapping from individuals to personalities or sets of attitudes or sets of beliefs, from thence to outcome mapping adjustments, and thus forms a separate stream of work. The ACT-based method is a mapping from the {\em evaluative beliefs of a group} to outcome mapping adjustments. It is thus much less invasive and more privacy preserving than personality or attitude tests.

The examples in Section~\ref{sec:sim} are quite simplistic, and uses outcome mappings that were arbitrarily selected. The point was only to show that biases in data based in affective stereotypes can be removed if both the stereotypes and the biases are measurable. This paper has presented one instance of such a dyad, but actually measuring and using this may present additional challenges. Further, some outcome mappings may still need to be adjusted by hand (i.e. with policy). This is because the attributes in question may not even be known to the decision makers, but are guaranteed by constitutional law and human rights.

\section{Conclusion}
The interplay between social capital (group) and autonomy (individual) is a foundational aspect of society~\cite{Fukuyama2014}.
As intelligent agents (so called AI) are becoming increasingly pervasive, two things happen. One, the nature of human work changes as increasingly people are either unemployed, or in very specialized jobs, such as maintaining robotic systems. 
Two, public policy decisions impacting people are increasingly made by these intelligent agents, or within parameters/models proposed by these intelligent agents. If the intelligent agents are not capable of understanding such an element of human existence as the balance between social capital and autonomy, 
then the models they propose for policy decisions will be heavily skewed towards one or the other (usually towards reducing autonomy through increased rules), and the popular reaction will appear random and uncorrelated. Relegating the variance in the popular reaction to noise in the system is a fundamental error in reasoning, as it is essentially projecting a two dimensional space of efficient behavior on a single dimension, and therefore misses the intersectional elements of it entirely. It is for this second reason, we believe, that models like ACT may be useful in helping people and politicians to better understand the complex world in which we live.

We stress again that the normalization we have proposed does not provably provide a solution to the intersectionality problem. It does, however, give a potential avenue for a novel solution concept, and we have attempted in this paper to show the basics of how such a solution could be found. The category labeling and sentiment measurements are currently limitations of the method, but this may be overcome with surveys done online, or using machine learning methods as noted.

As a final note, we stress the obvious importance of being careful with social engineering. We emphasize that our objective is for the algorithms proposed herein to be used strictly as informational devices for decision makers. A hiring decision could have an informed, unbiased reference point on which to base part of its decision, for example.

\bibliographystyle{ACM-Reference-Format}
\bibliography{refs}


\begin{thebibliography}{36}


\ifx \showCODEN    \undefined \def \showCODEN     #1{\unskip}     \fi
\ifx \showDOI      \undefined \def \showDOI       #1{#1}\fi
\ifx \showISBNx    \undefined \def \showISBNx     #1{\unskip}     \fi
\ifx \showISBNxiii \undefined \def \showISBNxiii  #1{\unskip}     \fi
\ifx \showISSN     \undefined \def \showISSN      #1{\unskip}     \fi
\ifx \showLCCN     \undefined \def \showLCCN      #1{\unskip}     \fi
\ifx \shownote     \undefined \def \shownote      #1{#1}          \fi
\ifx \showarticletitle \undefined \def \showarticletitle #1{#1}   \fi
\ifx \showURL      \undefined \def \showURL       {\relax}        \fi
\providecommand\bibfield[2]{#2}
\providecommand\bibinfo[2]{#2}
\providecommand\natexlab[1]{#1}
\providecommand\showeprint[2][]{arXiv:#2}

\bibitem[\protect\citeauthoryear{Bruch and Feinberg}{Bruch and
  Feinberg}{2017}]%
        {BruchFeinberg2017}
\bibfield{author}{\bibinfo{person}{Elizabeth Bruch} {and} \bibinfo{person}{Fred
  Feinberg}.} \bibinfo{year}{2017}\natexlab{}.
\newblock \showarticletitle{Decision-Making Processes in Social Contexts}.
\newblock \bibinfo{journal}{\emph{Annual Review of Sociology}}
  \bibinfo{volume}{43} (\bibinfo{year}{2017}), \bibinfo{pages}{207--227}.
\newblock


\bibitem[\protect\citeauthoryear{Buolamwini and Gebru}{Buolamwini and
  Gebru}{2018}]%
        {Buolamwini2018GenderSI}
\bibfield{author}{\bibinfo{person}{Joy Buolamwini} {and}
  \bibinfo{person}{Timnit Gebru}.} \bibinfo{year}{2018}\natexlab{}.
\newblock \showarticletitle{Gender Shades: Intersectional Accuracy Disparities
  in Commercial Gender Classification}. In \bibinfo{booktitle}{\emph{FAT}}.
\newblock


\bibitem[\protect\citeauthoryear{Durkheim}{Durkheim}{1893}]%
        {Durkheim1893}
\bibfield{author}{\bibinfo{person}{Emile Durkheim}.}
  \bibinfo{year}{2014/1893}\natexlab{}.
\newblock \bibinfo{booktitle}{\emph{The Division of Labor in Society}}.
\newblock \bibinfo{publisher}{Free Press}.
\newblock


\bibitem[\protect\citeauthoryear{Dwork, Hardt, Pitassi, Reingold, and
  Zemel}{Dwork et~al\mbox{.}}{2011}]%
        {Dwork2011}
\bibfield{author}{\bibinfo{person}{Cynthia Dwork}, \bibinfo{person}{Moritz
  Hardt}, \bibinfo{person}{Toniann Pitassi}, \bibinfo{person}{Omer Reingold},
  {and} \bibinfo{person}{Richard Zemel}.} \bibinfo{year}{2011}\natexlab{}.
\newblock \showarticletitle{Fairness through awareness}. In
  \bibinfo{booktitle}{\emph{Proceedings of Innovations of Theoretical Computer
  Science}}.
\newblock


\bibitem[\protect\citeauthoryear{Foulds, Islam, Keya, and Pan}{Foulds
  et~al\mbox{.}}{2019}]%
        {foulds2019intersectional}
\bibfield{author}{\bibinfo{person}{James Foulds}, \bibinfo{person}{Rashidul
  Islam}, \bibinfo{person}{Kamrun~Naher Keya}, {and} \bibinfo{person}{Shimei
  Pan}.} \bibinfo{year}{2019}\natexlab{}.
\newblock \bibinfo{title}{An Intersectional Definition of Fairness}.
\newblock
\newblock
\showeprint[arxiv]{1807.08362}~[cs.LG]


\bibitem[\protect\citeauthoryear{Francis and Heise}{Francis and Heise}{2006}]%
        {Francis2006}
\bibfield{author}{\bibinfo{person}{Clare Francis} {and}
  \bibinfo{person}{David~R. Heise}.} \bibinfo{year}{2006}\natexlab{}.
\newblock \bibinfo{title}{Mean Affective Ratings of 1,500 Concepts by Indiana
  University Undergraduates in 2002-3}.
\newblock \bibinfo{howpublished}{Computer file, Distributed at Affect Control
  Theory Website, Program Interact bayesact.ca}.
\newblock


\bibitem[\protect\citeauthoryear{Friston}{Friston}{2010}]%
        {Friston_brain_2010}
\bibfield{author}{\bibinfo{person}{Karl Friston}.}
  \bibinfo{year}{2010}\natexlab{}.
\newblock \showarticletitle{The free-energy principle: A unified brain theory?}
\newblock \bibinfo{journal}{\emph{Nature Reviews Neuroscience}}
  \bibinfo{volume}{11}, \bibinfo{number}{(2)} (\bibinfo{year}{2010}),
  \bibinfo{pages}{127--138}.
\newblock


\bibitem[\protect\citeauthoryear{Fukuyama}{Fukuyama}{2014}]%
        {Fukuyama2014}
\bibfield{author}{\bibinfo{person}{Francis Fukuyama}.}
  \bibinfo{year}{2014}\natexlab{}.
\newblock \bibinfo{booktitle}{\emph{Political Order and Political Decay}}.
\newblock \bibinfo{publisher}{Farrar, Strauss and Giroux}.
\newblock


\bibitem[\protect\citeauthoryear{Gollob}{Gollob}{1974}]%
        {Gollob1974}
\bibfield{author}{\bibinfo{person}{Harry~F. Gollob}.}
  \bibinfo{year}{1974}\natexlab{}.
\newblock \showarticletitle{The Subject-Verb-Object Approach to Social
  Cognition}.
\newblock \bibinfo{journal}{\emph{Psychological Review}}  \bibinfo{volume}{81}
  (\bibinfo{year}{1974}), \bibinfo{pages}{286--321}.
\newblock


\bibitem[\protect\citeauthoryear{Grimshaw and Ellinghaus}{Grimshaw and
  Ellinghaus}{1999}]%
        {Grimshaw1999}
\bibfield{author}{\bibinfo{person}{Patricia Grimshaw} {and}
  \bibinfo{person}{Katherine Ellinghaus}.} \bibinfo{year}{1999}\natexlab{}.
\newblock \showarticletitle{White women, Aboriginal women and the vote in
  Western Australia}.
\newblock \bibinfo{journal}{\emph{Studies in Western Australian History}}
  \bibinfo{volume}{19} (\bibinfo{year}{1999}), \bibinfo{pages}{1–--19}.
\newblock


\bibitem[\protect\citeauthoryear{Heise}{Heise}{2007}]%
        {Heise2007}
\bibfield{author}{\bibinfo{person}{David~R. Heise}.}
  \bibinfo{year}{2007}\natexlab{}.
\newblock \bibinfo{booktitle}{\emph{Expressive Order: Confirming Sentiments in
  Social Actions}}.
\newblock \bibinfo{publisher}{Springer}.
\newblock


\bibitem[\protect\citeauthoryear{Heise}{Heise}{2010}]%
        {Heise2010}
\bibfield{author}{\bibinfo{person}{David~R. Heise}.}
  \bibinfo{year}{2010}\natexlab{}.
\newblock \bibinfo{booktitle}{\emph{Surveying Cultures: Discovering Shared
  Conceptions and Sentiments}}.
\newblock \bibinfo{publisher}{Wiley}.
\newblock


\bibitem[\protect\citeauthoryear{Hoey}{Hoey}{2021}]%
        {Hoey2021b}
\bibfield{author}{\bibinfo{person}{Jesse Hoey}.}
  \bibinfo{year}{2021}\natexlab{}.
\newblock \showarticletitle{Freedom and Equality as Uncertainty in Groups}.
\newblock \bibinfo{journal}{\emph{Entropy}} \bibinfo{volume}{1384},
  \bibinfo{number}{23} (\bibinfo{year}{2021}).
\newblock


\bibitem[\protect\citeauthoryear{Hoey, MacKinnon, and Schr{\"o}der}{Hoey
  et~al\mbox{.}}{2021}]%
        {HoeyMacKinnon2020}
\bibfield{author}{\bibinfo{person}{Jesse Hoey}, \bibinfo{person}{Neil
  MacKinnon}, {and} \bibinfo{person}{Tobias Schr{\"o}der}.}
  \bibinfo{year}{2021}\natexlab{}.
\newblock \showarticletitle{Denotative and Connotative Control of Uncertainty:
  A Computational Dual-Process Model}.
\newblock \bibinfo{journal}{\emph{Judgment and Decision Making}}
  \bibinfo{volume}{16}, \bibinfo{number}{2} (\bibinfo{date}{March}
  \bibinfo{year}{2021}).
\newblock


\bibitem[\protect\citeauthoryear{Joseph and Morgan}{Joseph and Morgan}{2021}]%
        {JosephMorgan2021}
\bibfield{author}{\bibinfo{person}{Kenneth Joseph} {and}
  \bibinfo{person}{Jonathan~H. Morgan}.} \bibinfo{year}{2021}\natexlab{}.
\newblock \showarticletitle{Friend or Foe: A Review and Synthesis of
  Computational Models of the Identity Labeling Problem}.
\newblock \bibinfo{journal}{\emph{The Journal of Mathematical Sociology}}
  \bibinfo{volume}{0}, \bibinfo{number}{0} (\bibinfo{year}{2021}),
  \bibinfo{pages}{1--35}.
\newblock
\urldef\tempurl%
\url{https://doi.org/10.1080/0022250X.2021.1923016}
\showDOI{\tempurl}


\bibitem[\protect\citeauthoryear{Joseph, Wei, Benigni, and Carley}{Joseph
  et~al\mbox{.}}{2016}]%
        {Joseph2016}
\bibfield{author}{\bibinfo{person}{Kenneth Joseph}, \bibinfo{person}{Wei Wei},
  \bibinfo{person}{Matthew Benigni}, {and} \bibinfo{person}{Kathleen~M.
  Carley}.} \bibinfo{year}{2016}\natexlab{}.
\newblock \showarticletitle{A social-event based approach to sentiment analysis
  of identities and behaviors in text}.
\newblock \bibinfo{journal}{\emph{The Journal of Mathematical Sociology}}
  \bibinfo{volume}{40}, \bibinfo{number}{3} (\bibinfo{year}{2016}),
  \bibinfo{pages}{137--166}.
\newblock
\urldef\tempurl%
\url{https://doi.org/10.1080/0022250X.2016.1159206}
\showDOI{\tempurl}


\bibitem[\protect\citeauthoryear{Kong}{Kong}{2021}]%
        {Kong2021}
\bibfield{author}{\bibinfo{person}{Youjin Kong}.}
  \bibinfo{year}{2021}\natexlab{}.
\newblock \showarticletitle{Intersectional Fairness in {AI}? A Critical
  Analysis}. In \bibinfo{booktitle}{\emph{Conference on Feminism, Social
  Justice, and AI}}. \bibinfo{address}{Waterloo, Canada (online)}.
\newblock


\bibitem[\protect\citeauthoryear{Kouloumpis, Wilson, and Moore}{Kouloumpis
  et~al\mbox{.}}{2011}]%
        {kouloumpis2011}
\bibfield{author}{\bibinfo{person}{Efthymios Kouloumpis},
  \bibinfo{person}{Theresa Wilson}, {and} \bibinfo{person}{Johanna Moore}.}
  \bibinfo{year}{2011}\natexlab{}.
\newblock \showarticletitle{Twitter sentiment analysis: The good the bad and
  the {OMG}!}. In \bibinfo{booktitle}{\emph{Fifth International AAAI conference
  on weblogs and social media}}.
\newblock


\bibitem[\protect\citeauthoryear{Kozlowski, Taddy, and Evans}{Kozlowski
  et~al\mbox{.}}{2019}]%
        {kozlowski2019geometry}
\bibfield{author}{\bibinfo{person}{Austin~C. Kozlowski}, \bibinfo{person}{Matt
  Taddy}, {and} \bibinfo{person}{James~A. Evans}.}
  \bibinfo{year}{2019}\natexlab{}.
\newblock \showarticletitle{The Geometry of Culture: Analyzing the Meanings of
  Class through Word Embeddings}.
\newblock \bibinfo{journal}{\emph{American Sociological Review}}
  \bibinfo{volume}{84}, \bibinfo{number}{5} (\bibinfo{year}{2019}),
  \bibinfo{pages}{905--949}.
\newblock
\urldef\tempurl%
\url{https://doi.org/10.1177/0003122419877135}
\showDOI{\tempurl}


\bibitem[\protect\citeauthoryear{Lepori}{Lepori}{2020}]%
        {lepori2020unequal}
\bibfield{author}{\bibinfo{person}{Michael~A. Lepori}.}
  \bibinfo{year}{2020}\natexlab{}.
\newblock \bibinfo{title}{Unequal Representations: Analyzing Intersectional
  Biases in Word Embeddings Using Representational Similarity Analysis}.
\newblock
\newblock
\showeprint[arxiv]{2011.12086}~[cs.CL]


\bibitem[\protect\citeauthoryear{Lewis, Ghandeharioun, Fedor, Pedrelli, Picard,
  and Mischoulon}{Lewis et~al\mbox{.}}{2021}]%
        {Lewis2021}
\bibfield{author}{\bibinfo{person}{Rob Lewis}, \bibinfo{person}{Asma
  Ghandeharioun}, \bibinfo{person}{Szymon Fedor}, \bibinfo{person}{Paola
  Pedrelli}, \bibinfo{person}{Rosalind Picard}, {and} \bibinfo{person}{David
  Mischoulon}.} \bibinfo{year}{2021}\natexlab{}.
\newblock \showarticletitle{Mixed Effects Random Forests for Personalised
  Predictions of Clinical Depression Severity}. In
  \bibinfo{booktitle}{\emph{ICML 2021 Computational Approaches to Mental Health
  workshop}}.
\newblock


\bibitem[\protect\citeauthoryear{MacKinnon and Heise}{MacKinnon and
  Heise}{2010}]%
        {MacKinnonHeise2010}
\bibfield{author}{\bibinfo{person}{Neil~J. MacKinnon} {and}
  \bibinfo{person}{David~R. Heise}.} \bibinfo{year}{2010}\natexlab{}.
\newblock \bibinfo{booktitle}{\emph{Self, identity and social institutions}}.
\newblock \bibinfo{publisher}{Palgrave and Macmillan}, \bibinfo{address}{New
  York, NY}.
\newblock


\bibitem[\protect\citeauthoryear{Mead}{Mead}{1934}]%
        {Mead1934}
\bibfield{author}{\bibinfo{person}{George~Herbert Mead}.}
  \bibinfo{year}{1934}\natexlab{}.
\newblock \bibinfo{booktitle}{\emph{Mind, Self and Society}}.
\newblock \bibinfo{publisher}{University of Chicago Press}.
\newblock


\bibitem[\protect\citeauthoryear{Nichols and Stahl}{Nichols and Stahl}{2019}]%
        {Nichols2019}
\bibfield{author}{\bibinfo{person}{Sue Nichols} {and} \bibinfo{person}{Garth
  Stahl}.} \bibinfo{year}{2019}\natexlab{}.
\newblock \showarticletitle{Intersectionality in higher education research: a
  systematic literature review}.
\newblock \bibinfo{journal}{\emph{Higher Education Research and Development}}
  \bibinfo{volume}{38} (\bibinfo{date}{07} \bibinfo{year}{2019}),
  \bibinfo{pages}{1--14}.
\newblock
\urldef\tempurl%
\url{https://doi.org/10.1080/07294360.2019.1638348}
\showDOI{\tempurl}


\bibitem[\protect\citeauthoryear{Osgood, Suci, and Tannenbaum}{Osgood
  et~al\mbox{.}}{1957}]%
        {Osgood1957}
\bibfield{author}{\bibinfo{person}{Charles~E. Osgood}, \bibinfo{person}{G.~J.
  Suci}, {and} \bibinfo{person}{Percy~H. Tannenbaum}.}
  \bibinfo{year}{1957}\natexlab{}.
\newblock \bibinfo{booktitle}{\emph{The Measurement of Meaning}}.
\newblock \bibinfo{publisher}{University of Illinois Press},
  \bibinfo{address}{Urbana}.
\newblock


\bibitem[\protect\citeauthoryear{Page}{Page}{2007}]%
        {Page2007}
\bibfield{author}{\bibinfo{person}{Scott~E. Page}.}
  \bibinfo{year}{2007}\natexlab{}.
\newblock \bibinfo{booktitle}{\emph{The Difference: How the power of diversity
  creates better groups, firms, schools and societies}}.
\newblock \bibinfo{publisher}{Princeton University Press}.
\newblock


\bibitem[\protect\citeauthoryear{Pearl}{Pearl}{1988}]%
        {Pearl88}
\bibfield{author}{\bibinfo{person}{Judea Pearl}.}
  \bibinfo{year}{1988}\natexlab{}.
\newblock \bibinfo{booktitle}{\emph{Probabilistic Reasoning in Intelligent
  Systems: Networks of Plausible Inference.}}
\newblock \bibinfo{publisher}{Morgan Kaufmann}, \bibinfo{address}{San Mateo,
  CA}.
\newblock


\bibitem[\protect\citeauthoryear{Powers}{Powers}{1973}]%
        {Powers1973}
\bibfield{author}{\bibinfo{person}{William~T. Powers}.}
  \bibinfo{year}{1973}\natexlab{}.
\newblock \bibinfo{booktitle}{\emph{Behavior: The control of perception}}.
\newblock \bibinfo{publisher}{Aldine publishing co.},
  \bibinfo{address}{Chicago}.
\newblock


\bibitem[\protect\citeauthoryear{Redhead and Power}{Redhead and Power}{2021}]%
        {RedheadPower2021}
\bibfield{author}{\bibinfo{person}{Daniel Redhead} {and}
  \bibinfo{person}{Eleanor~A. Power}.} \bibinfo{year}{2021}\natexlab{}.
\newblock \showarticletitle{Social Hierarchies and Social Networks in Humans}.
\newblock \bibinfo{journal}{\emph{Philosophical Transactions of the Royal
  Society B: Biological Sciences}} (\bibinfo{year}{2021}).
\newblock
\urldef\tempurl%
\url{https://doi.org/10.1098/rstb.2020.0440}
\showDOI{\tempurl}


\bibitem[\protect\citeauthoryear{Rozado}{Rozado}{2020}]%
        {Rozado2020}
\bibfield{author}{\bibinfo{person}{David Rozado}.}
  \bibinfo{year}{2020}\natexlab{}.
\newblock \showarticletitle{Wide range screening of algorithmic bias in word
  embedding models using large sentiment lexicons reveals underreported bias
  types}.
\newblock \bibinfo{journal}{\emph{PloS one}}  \bibinfo{volume}{15}
  (\bibinfo{year}{2020}).
\newblock
\urldef\tempurl%
\url{https://doi.org/10.1371/journal.pone.0231189}
\showDOI{\tempurl}


\bibitem[\protect\citeauthoryear{Scotti}{Scotti}{2017}]%
        {Scotti2017}
\bibfield{author}{\bibinfo{person}{Valentina~Rita Scotti}.}
  \bibinfo{year}{2017}\natexlab{}.
\newblock \showarticletitle{Women’s rights and minorities’ rights in
  Canada. The challenges of intersectionality in Supreme Court jurisprudence}.
\newblock \bibinfo{journal}{\emph{Perspectives on Federalism}}
  \bibinfo{volume}{9} (\bibinfo{year}{2017}).
\newblock


\bibitem[\protect\citeauthoryear{Stark and Hoey}{Stark and Hoey}{2020}]%
        {StarkHoey2020}
\bibfield{author}{\bibinfo{person}{Luke Stark} {and} \bibinfo{person}{Jesse
  Hoey}.} \bibinfo{year}{2020}\natexlab{}.
\newblock \showarticletitle{The Ethics of Emotion in {AI} Systems}. In
  \bibinfo{booktitle}{\emph{Proc. FaCCT*}}. \bibinfo{publisher}{ACM},
  \bibinfo{pages}{782--793}.
\newblock
\urldef\tempurl%
\url{https://doi.org/10.1145/3442188.3445939}
\showDOI{\tempurl}


\bibitem[\protect\citeauthoryear{Taleb}{Taleb}{2012}]%
        {Taleb2012}
\bibfield{author}{\bibinfo{person}{Nassim Taleb}.}
  \bibinfo{year}{2012}\natexlab{}.
\newblock \bibinfo{booktitle}{\emph{Antifragile: Things That Gain Through
  Disorder.}}
\newblock \bibinfo{publisher}{Random House}.
\newblock


\bibitem[\protect\citeauthoryear{Turan, Elafros, Logie, Banik, Turan, Crockett,
  Pescosolido, and Murray}{Turan et~al\mbox{.}}{2019}]%
        {Turan2019}
\bibfield{author}{\bibinfo{person}{Janet~M. Turan}, \bibinfo{person}{Melissa~A.
  Elafros}, \bibinfo{person}{Carmen~H. Logie}, \bibinfo{person}{Swagata Banik},
  \bibinfo{person}{Bulent Turan}, \bibinfo{person}{Kaylee~B. Crockett},
  \bibinfo{person}{Bernice Pescosolido}, {and} \bibinfo{person}{Sarah~M.
  Murray}.} \bibinfo{year}{2019}\natexlab{}.
\newblock \showarticletitle{Challenges and opportunities in examining and
  addressing intersectional stigma and health}.
\newblock \bibinfo{journal}{\emph{BMC Medicine}} \bibinfo{volume}{17},
  \bibinfo{number}{7} (\bibinfo{year}{2019}).
\newblock


\bibitem[\protect\citeauthoryear{{Van Loon} and Freese}{{Van Loon} and
  Freese}{2021}]%
        {VanLoon2022}
\bibfield{author}{\bibinfo{person}{Austin {Van Loon}} {and}
  \bibinfo{person}{Jeremy Freese}.} \bibinfo{year}{2021}\natexlab{}.
\newblock \showarticletitle{Word embeddings reveal how fundamental sentiments
  structure natural language}.
\newblock \bibinfo{journal}{\emph{American Behavioral Scientist}}
  (\bibinfo{year}{2021}).
\newblock
\newblock
\shownote{In Press.}


\bibitem[\protect\citeauthoryear{Wellens}{Wellens}{2008}]%
        {Wellens2008}
\bibfield{author}{\bibinfo{person}{Pieter Wellens}.}
  \bibinfo{year}{2008}\natexlab{}.
\newblock \showarticletitle{Coping with Combinatorial Uncertainty in Word
  Learning: A Flexible Usage-Based Mode}. In \bibinfo{booktitle}{\emph{7th
  International Conference on The Evolution of Language}}
  \emph{(\bibinfo{series}{7th International Conference on The Evolution of
  Language})}, \bibfield{editor}{\bibinfo{person}{A.d.m Smith},
  \bibinfo{person}{K.~Smith}, {and} \bibinfo{person}{{R.f. I} Cancho}} (Eds.).
  \bibinfo{publisher}{World Scientific Publishing}, \bibinfo{pages}{370--377}.
\newblock
\showISBNx{981-277-611-7}
\newblock
\shownote{A.D.M Smith, K. Smith, R.F. i Cancho.}


\end{thebibliography}

\end{document}